# Ultrafast nonadiabatic dynamics of tetraphenylsubstituted nitrogen-based heterocycles

June 19, 2026

Javier Hernández-Rodríguez Alberto Martín Santa Daría Susana Gómez-Carrasco Sandra Gómez*
J. Hernández-Rodríguez, Prof. A. Martín Santa Daría, Prof. S. Gómez-Carrasco, Prof. S. Gómez
Departamento de Química Física, Universidad de Salamanca https://ror.org/02f40zc51, 37008, Spain

Prof. S. Gómez
Departamento de Química, Módulo 13, Universidad Autónoma de Madrid https://ror.org/01cby8j38, 28049, Spain
Email Address: sandra.gomezr@uam.es

**Abstract**

Tetraphenylpyrazine (TPP) and 2,3,4,5-tetraphenyl-1H-pyrrole (TePP) are closely related heterocycles bearing four phenyl substituents, whose structural similarity makes them a useful pair for comparing how intramolecular flexibility influences excited-state relaxation and emission in the gas phase and in the solid state. TPP is a prototypical solid-state luminescence enhancement (SLE) emitter, exhibiting a markedly increased quantum yield upon molecular aggregation. In contrast, TePP displays similar quantum yields in solution and solid state, characteristic of dual-state emission (DSE). This behaviour indicates that intramolecular rotations are already significantly hindered in the isolated-molecule regime, consistent with our previous observations for TPP and other solid-state emitters (Hernández-Rodríguez et al., ChemPhysChem, 2024, 25, e202400563). To unravel the excited-state dynamics underlying this contrasting behaviour, we performed mixed quantum–classical trajectory simulations on a single molecule of TPP and TePP employing the surface-hopping method. Twelve singlet states were included at the TD-B3LYP-D3/def2-SVP level, which were previously benchmarked against coupled cluster methods. Simulated observables such as gas phase ultrafast electron diffraction (GUED) and time-resolved fluorescence (TR-FL) signals allow us to dissect the distinct deactivation pathways operating in both systems in the gas phase, while also providing mechanistic insight into how these pathways are expected to evolve in solution and solid-state environments.

## 1 Introduction

Fluorescent organic chromophores are widely used in areas such as cellular imaging and organic light-emitting diodes (OLEDs),[1, 2, 3, 4] where performance depends on how their photophysical properties change between dilute solution, gas phase, and the solid state. Traditionally, many organic emitters suffer from aggregation-caused quenching (ACQ), which severely reduces their efficiency in condensed phases and limits their usefulness in optoelectronic applications.[5, 6]

In response to this limitation, efforts have focused on molecules that show solid-state luminescence enhancement (SLE), systems whose emission becomes significantly more intense upon aggregation or crystallization compared to solution.[7, 8, 9, 10] Within this broad SLE family, luminogens such as tetraphenylethylene (TPE) and silole derivatives have been exploited in optoelectronics and photodynamic therapy thanks to their combination of bright solid-state emission and efficient generation of reactive oxygen species. [11, 12, 13, 14, 15, 16, 17]

In contrast, dual-state emission (DSE) chromophores maintain efficient fluorescence in both solution and the solid state, bridging the gap between purely solution-bright and purely solid-bright materials and enabling applications where stable emission across different environments is required. Recent reviews show that DSE luminogens are particularly promising for sensing or bioimaging although their biomedical impact remains more limited than SLE photosensitizers. [18, 19, 20, 21]

To understand both SLE and DSE, it is important to consider the effect produced by Restricted Access to Conical Intersections (RACI),[24] which describes how aggregation or crystallisation limits key nuclear motion that would otherwise lead the system towards a conical intersection (CI) $S_1/S_0$ and non-radiative decay.[25, 26, 27] This converts weakly emissive molecules in solution into strong emitters in the solid state, i.e., producing SLE rather than ACQ.

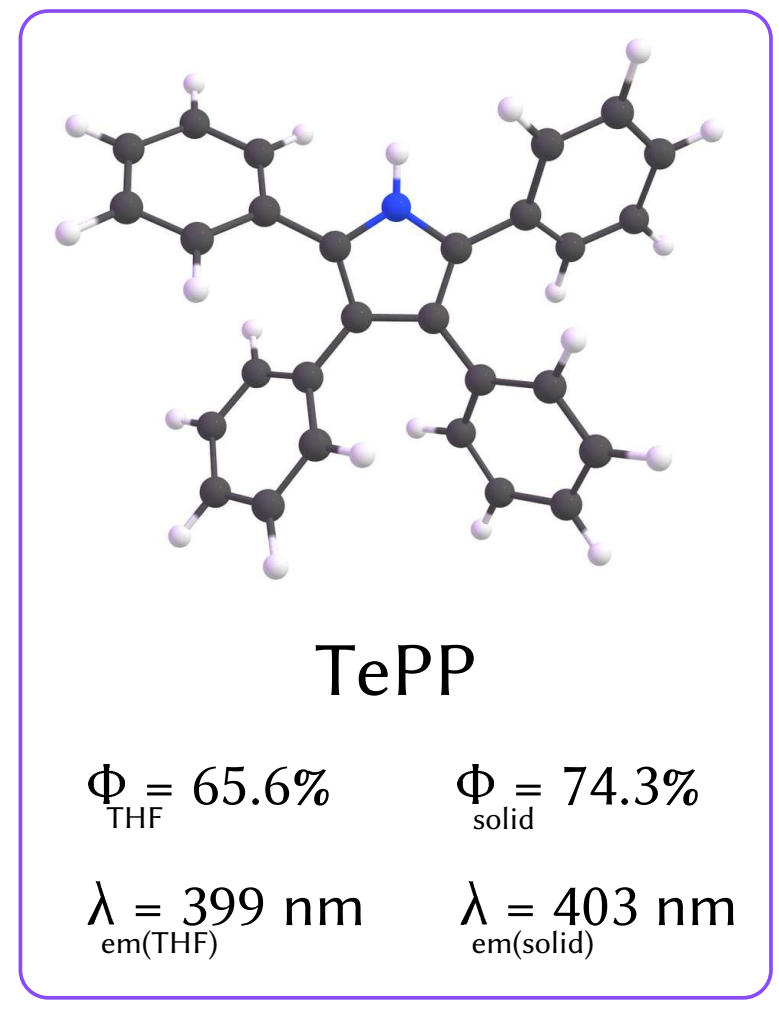

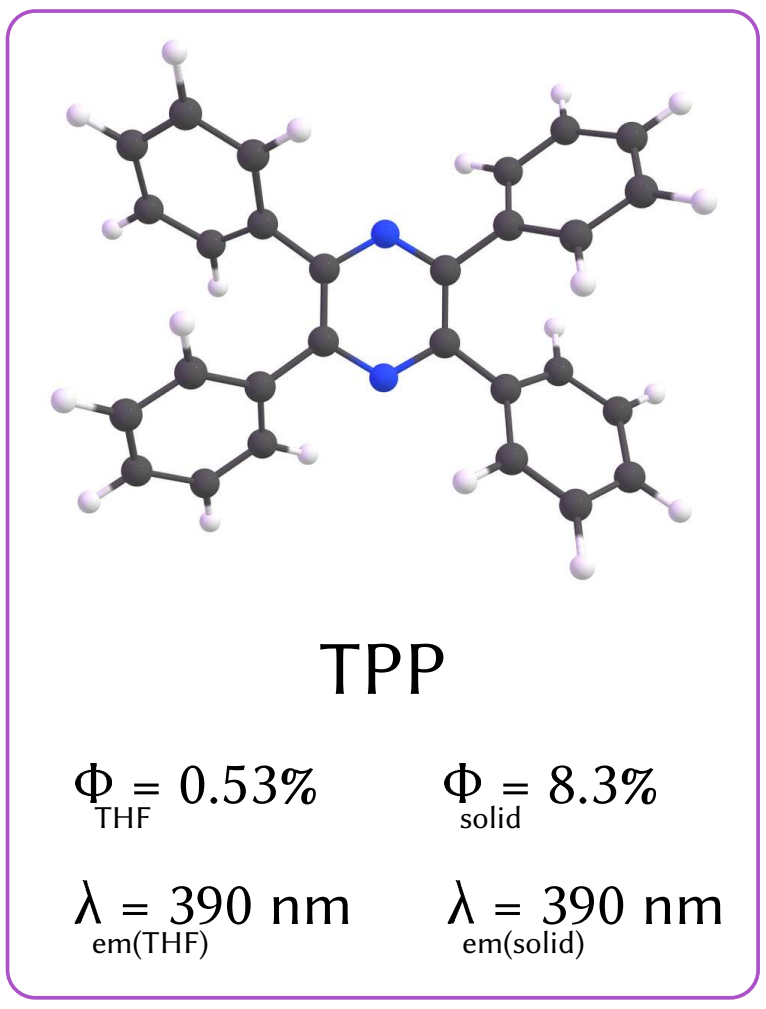

Figure 1: Optimised molecular structures (B3LYP/6-31G**, computed in Gaussian) of TePP and TPP with their experimental optical properties extracted from Refs. 22 and 23: quantum yields and emission wavelengths in solid and liquid phases. TePP: 399 nm (3.11 eV) in THF and 403 nm (3.08 eV) in the solid state; TPP: 390 nm (3.18 eV) in both phases. The excitation energies were 317 nm (3.91 eV) for TePP and 338 nm (3.67 eV) for TPP.

DSE luminogens are in contrast designed so the access to these conical intersection is already unfavorable in solution, through intrinsic rigidity or restricted twists, and, at the same time, their aggregates avoid ACQ by suppressing detrimental $\pi - \pi$ stacking.[28]

In addition, the Restriction of Intramolecular Motions (RIM) model has also been considered to describe SLE mechanisms. Operating under the harmonic approximation, it attributes the lack of emission in solution to the dissipation of electronic energy into low-frequency vibrational modes that act as energy acceptors near the equilibrium geometry, which limits its applicability to systems involving only small geometry changes. By contrast, the RACI model describes how large-scale nuclear distortions, such as double-bond torsion or ring puckering, lead to a conical intersection in solution that is inaccessible in the aggregate phase because of steric effects.[27, 29, 30] Consequently, while RIM is effective for quantitative treatments of lifetimes in the aggregate phase, where harmonicity often holds, the RACI model is essential for identifying the specific non-radiative deactivation coordinates responsible for quenching in an unconstrained solution environment.

In this work, the focus is placed on two tetraphenyl-substituted nitrogen heterocycles displayed in Fig.1: tetraphenylpyrazine (TPP), which exhibits pronounced solid-state luminescence enhancement, and 2,3,4,5-tetraphenyl-1H-pyrrole (TePP), a dual-state emitter with comparable quantum yields in solution and solid phases. By comparing their excited-state dynamics, this study aims to discover how intrinsic non-radiative pathways differ between compounds that undergo SLE and DSE in the gas phase, without taking into account molecular aggregation or environment.

Tetraphenylpyrazine (TPP) is a rigid heterocycle consisting on a central pyrazine core with four peripheral phenyl rings, giving rise to a highly twisted $\pi$-conjugated framework. In solution, TPP displays relatively weak blue emission, whereas in the solid state its quantum yield increases considerably. Besides, both the emission efficiency and wavelength can be tuned by peripheral substitution, making TPP derivatives robust solid-state emitters for OLEDs, sensors based on metallic organic frameworks (MOFs) or bioimaging.[23, 31, 32, 33]

By contrast, 2,3,4,5-tetraphenyl-1H-pyrrole (TePP) combines a pyrrolic core with four phenyl arranged around the ring, leading to a similarly but electronically distinct architecture as explained in Ref.22. This is because of the different function of the substituents, 2,5-phenyl rings extend the conjugation, locking the conformation and increasing the rigidity, while 3,4-phenyls adopt twisted conformation limiting the close packing. [28, 34] TePP can be obtained through a simple one-step synthesis and exhibits balanced blue emission with comparable quantum yields in dilute solution and in the solid state, which has established it as a prototypical dual-state emitter and as a useful building block for more complex pyrrole-based luminescent materials.[35, 36]

Beyond their contrasting emissive behavior, TPP and TePP can serve as representative systems to explore how nonadiabatic dynamics shape the photophysics of rigid, tetraphenyl-substituted chromophores. Nonadiabatic relaxation in this class of molecules typically involves ultrafast population transfer among multiple singlet states mediated by conical intersections and strongly coupled vibrational modes, requiring methods that can track coupled electronic–nuclear motion in real time.

Mixed quantum–classical trajectory approaches, in particular Tully's fewest-switches surface hopping (TSH),[37, 38] have been widely used for studying such photoinduced processes in organic emitters because they allow on-the-fly treatment of several excited states at a feasible computational cost. For pyrazine-like and related conjugated systems, TSH and wavepacket-based schemes such as MCTDH using linear vibronic couplings have been successfully used to characterize internal conversion pathways, validate diabatic Hamiltonians, and rationalize ultrafast spectroscopic signals.[39, 40]

Our recent work, provided a detailed electronic-structure characterization of tetraphenylpyrazine and three related donor–acceptor photosensitizers. [41] It showed that TPP features a dense manifold of bright and dark states, with the lowest band arising from nearly degenerate excitations whose energies are strongly modulated by phenyl torsions, suggesting that internal rotations and dark states are relevant to its solid-state luminescence enhancement.

This electronic-structure picture motivates the present work, where nonadiabatic dynamics are used to track how population evolves among these states after photoexcitation, propagating ensembles of trajectories initiated from Wigner-sampled ground-state distributions and excited into the lowest singlet manifold. This strategy enables a direct, state-resolved comparison of the excited-state population, geometric relaxation, and simulated experimental observables of TPP and TePP by linking their intrinsic gas phase nonadiabatic dynamics to their solid-state luminescence enhancement and dual-state emission behavior in condense phases.

This work is also contributing to the ongoing effort to establish reliable benchmarks in nonadiabatic molecular dynamics, trying to address a gap in theoretical photodynamics research.[42, 43]

The manuscript is divided as follows: In Section 2, the employed methodologies for electronic structure and mixed quantum-classical nonadiabatic dynamics are presented. In Section 3, the static results are first presented and afterwards, the dynamics results, including state populations, geometrical changes and simulated experimental spectra. In Section 4, the conclusions of the work are summarized.

# 2 Methods and theoretical description

## 2.1 Electronic structure details

The level of theory for the electronic structure calculations was selected based on Ref. 41, where a benchmark of several TDDFT functionals and basis sets was carried out for tetraphenylpyrazine (TPP). In that work, the performance of different TDDFT approaches was evaluated against Equation of Motion EOM-CCSD reference data and against the experimental UV-Vis absorption spectrum reported by M. Chen *et al.*[23].

The best compromise between accuracy and computational cost was achieved with the TD-B3LYP/def2-SVP method. This level of theory was therefore used for computing vertical excitation energies and for performing the on-the-fly nonadiabatic dynamics using the ORCA 5.0 electronic structure package. [44, 45] The vibrational frequencies of the TPP ground state were obtained with Gaussian 16 [46] at the equivalent B3LYP/6-31G** level of theory. B3LYP and CCSD(T) were paired with Pople-type and correlation-consistent basis sets respectively, consistent with those used in their original development. For B3LYP, the more modern Ahlrichs-type def2-SVP basis set (commonly used with ORCA) was also employed, yielding slightly improved agreement with experiment.

Although no separate benchmark study was performed for tetraphenylpyrrol (TePP), we adopted the same TD-B3LYP-based methodology for this molecule in order to enable a direct and consistent comparison with TPP. Vertical excitation energies were initially evaluated at the EOM- CCSD/cc-pVDZ level of theory to provide a high-level reference for the electronic transitions. However, due to the prohibitive computational cost associated with solving the left-hand side EOM equations for properties in a canonical framework, oscillator strengths (f) were computed for the TePP system using the Density Fitting Local Coupled Cluster Singles and Doubles (DF-LCC2-LR)[47, 48] method as implemented in MOLPRO package.[49] The frontier orbitals that characterize the low-lying electronic states of both systems at the TD-B3LYP level are shown in Figure 2.

## 2.2 Mixed quantum-classical nonadiabatic dynamics: the Tully Surface Hopping method

In the Tully surface hopping (TSH) approach [37, 38], the nuclear degrees of freedom are treated classically and propagated according to Newton's equations of motion on a single adiabatic potential energy surface $V_i(\mathbf{R}(t))$. At any given time, each atom $A$ evolves under the force given by the gradient of the active electronic state, and the nuclear positions $\mathbf{R}_A(t)$ and velocities $\mathbf{v}_A(t)$ are integrated using the velocity-Verlet algorithm [50].

The nuclei therefore move on a single potential energy surface associated with the current adiabatic electronic state. In regions where the nonadiabatic or spin-orbit coupling between electronic states becomes significant, transitions between adiabatic states are allowed through stochastic "hops", which give the method its name. These hops enable the system to change the active electronic state while maintaining a classical description of the nuclear motion.

The probability of a hop between electronic states depends on the time-dependent electronic state coefficients, which are propagated alongside the nuclear dynamics by solving the time-dependent Schrödinger equation in the adiabatic representation:

$$\frac{\mathrm{d}c_\beta(t)}{\mathrm{d}t} = -\sum_\alpha \left[ \mathrm{i} \underbrace{\left\langle \Psi_\beta^{\mathrm{el}} \middle| \hat{H}^{\mathrm{el}} \middle| \Psi_\alpha^{\mathrm{el}} \right\rangle}_{H_{\beta\alpha}} + \underbrace{\left\langle \Psi_\beta^{\mathrm{el}} \middle| \frac{\mathrm{d}}{\mathrm{d}t} \middle| \Psi_\alpha^{\mathrm{el}} \right\rangle}_{K_{\beta\alpha}} \right] c_\alpha(t). \tag{1}$$

The Hamiltonian term is directly calculated via electronic structure methods and the second term can be calculated from the nonadiabatic coupling between electronic states and velocities:
$K_{\beta\alpha} = \left\langle \Psi_\beta^{\mathrm{el}} \middle| \nabla_R \middle| \Psi_\alpha^{\mathrm{el}} \right\rangle \mathbf{v}_R$.

Since individual surface-hopping trajectories evolve on a single adiabatic electronic state, an ensemble of trajectories is required in order to mimic the spatial and momentum spread of a quantum nuclear wavepacket. In this work, the initial nuclear positions and momenta were sampled from a Wigner distribution, allowing the ensemble to represent initial vibrational phase-space correlations.

In the TSH implementation in SHARC[51] used in this work, whether a hop is performed from one state to another is determined by the ant-eater technique, generating a random number between 0 and 1 and comparing the number with the hopping probabilities given by the hopping algorithm, which is a function of the old and new electronic coefficients of the involved states and the propagator matrix. The kinetic energy is adjusted after a hop by rescaling the nuclear velocities so that it equals $\mathrm{E}_{tot}-\mathrm{E}_{pot}$. A hop is frustrated if the potential energy on the new state would exceed the total energy.

In surface hopping, each trajectory evolves on a single adiabatic electronic state and follows its corresponding nuclear gradient, while simultaneously carrying electronic coefficients for all other states. Because all electronic components are propagated along the same nuclear path, coherence between states can be artificially preserved. When two electronic states become close in energy, their coefficients may change and allow population transfer even in the absence of a hop, leading to spurious overcoherence and unphysical interference effects. To correct for this artifact, decoherence schemes are commonly applied. In this work, we employ the energy-based decoherence correction of Granucci and Persico, which exponentially damps the amplitudes of non-active electronic states as a function of the kinetic energy and the energy gap between states [52].

The electronic populations displayed here correspond to "quantum populations", i.e., the square of the electronic coefficient of each electronic state. The populations have been transformed to the diabatic picture using the matrix product of the reference overlap matrix, which is calculated at the beginning of the propagation and tags every state at the Franck-Condon region (every Wigner initial condition) with respect to the vertical energies at the optimised geometries.

The nonadiabatic dynamics of tetraphenylpyrazine (TPP) and tetraphenylpyrrol (TePP) were simulated using the surface-hopping trajectory method as implemented in SHARC. All nonadiabatic dynamics were performed without imposing any symmetry constraints. A total of 214 trajectories were initially propagated for TPP and 159 for TePP. For each system, the initial excitation was distributed among the $S_1$-$S_3$ states according to the relative oscillator strengths at the Franck–Condon geometry. The resulting population ratios were 0.58:0.42:0.00 for TPP and 0.65:0.26:0.09 for TePP. Energy conservation criteria were enforced throughout the dynamics: trajectories were discarded if the total energy drift exceeded 0.3 eV over the full simulation or 0.2 eV between consecutive steps. After applying these thresholds, 206 trajectories were retained for TPP and 148 for TePP, and these were used for the subsequent statistical analysis. Convergence of the electronic populations with respect to the number of trajectories was explicitly evaluated for both systems by analyzing two different subsets of increasing ensemble size, demonstrating that the number of trajectories employed is sufficient; the corresponding convergence analysis is presented in section S2 of the Supporting Information.

# 3 Results and discussion

## 3.1 Electronic character of the states at the Frank-Condon region

At the Franck-Condon geometry, the excited-state manifolds of TePP and TPP exhibit markedly different diabatic characters, reflecting their distinct donor-acceptor architectures and molecular symmetries. In TePP, all bright states arise from a single electronic excitation originating from a pyrrole-centered $\pi$ orbital. The $S_1$ state corresponds to the HOMO (Orb.098)→LUMO (Orb.099) transition, in which the acceptor orbital is localized on the two C–C single bonds that connect the pyrrole to the phenyl rings adjacent to the nitrogen atom. This transition enhances partial double-bond character at these linkages and extends conjugation into the corresponding phenyl rings. The $S_2$ state preserves the same donor but promotes the electron into the LUMO+1 (Orb.100) orbital, which is delocalized over all four phenyl substituents. In this diabat, the phenyl rings opposite the nitrogen exhibit bonding character, while the two rings adjacent to the nitrogen show antibonding density. The next two higher diabats ($S_3$ and $S_4$) share the same pyrrole-donor character but differ in that the electron is transferred into orbitals that are $\pi$-antibonding (Orb.102 and Orb.103) across the four phenyl rings. None of these acceptor orbitals retain electron density on the pyrrole fragment, so all low-lying TePP excitations possess intramolecular charge-transfer (CT) character.

Figure 2: Orbitals involved in the main excitations of TePP (top) and TPP (bottom) computed at the TD-B3LYP/def2-SVP level of theory. Orbitals are shown at the respective optimized ground-state geometries, with isosurfaces plotted at $\pm 0.045$ a.u.

To see the excited state assigned, see table 1.

The electronic structure of TPP produces a manifold of locally excited (LE) $\pi\pi^*$ states on the pyrazine core, accompanied by higher-lying $n\pi^*$ excitations. The $S_1$ and $S_2$ diabats both originate from a pyrazine-centered $\pi$-bonding orbital (Orb.101) but differ in their acceptor orbitals: $S_1$ is the HOMO (Orb.101) $n\pi^*$ (Orb.103) excitation localized on the pyrazine ring, whereas $S_2$ populates an orbital that displays double-bond character along the conjugated paths connecting the central heterocycle to the four phenyl groups (Orb.102). Because the four phenyl substituents in TPP are symmetry-equivalent at the Franck-Condon geometry, the electron density in $S_2$ is distributed more uniformly than in TePP. Higher in the manifold, the $S_3$ and $S_4$ diabats correspond to $n\pi^*$ excitations from the nitrogen lone pairs on the pyrazine (Orb. 099 and Orb.100), promoting electrons either into the pyrazine-centered $\pi^*$ orbital (Orb.103) or into the orbital associated with enhanced pyrazine-phenyl double-bond character (Orb.102). Thus, unlike TePP, the lowest excited states of TPP are non–CT, largely delocalized over the molecular framework, and distinguished only at higher energies by localized $n\pi^*$ contributions.

The computed absorption spectra of TePP and TPP are shown in Fig. 3, where the spectrum obtained from a Wigner distribution is compared against the experimental one. For TePP, the lowest energy band is dominated by the bright 1B state in good agreement with the experimental absorption peak at 3.80 eV. The higher-lying states contribute to a shoulder at higher energies, consistent with the dense manifold of CT states described above. For TPP, the two nearly degenerate bright states 1B and 1A give rise to a band centered at 3.63 eV. The overall agreement between TD-B3LYP/def2-SVP and the experimental spectra, supports the electronic structure description adopted for the non-adiabatic dynamics.

Table 1: Energies (eV), oscillator strengths (in parentheses), dominant orbital excitations, and electronic state symmetries of the low-lying excited states of TePP (top) and TPP (bottom) at the Franck-Condon geometry. Results are reported assuming the $C_2$ point group and were obtained using TD-B3LYP and coupled-cluster-based methods.

| State | Symm. /Excitation | TD-B3LYP | EOM-CCSD | DF-LCC2-LR |
|---|---|---|---|---|
| | | TePP | | |
| $S_1$ | 1B (098 $\rightarrow$ 099) | 3.98 (0.53) | 4.71 | 4.41 (0.58) |
| $S_2$ | 1A (098 $\rightarrow$ 100) | 4.12 (0.10) | 4.88 | 4.65 (0.05) |
| $S_3$ | 2A (098 $\rightarrow$ 101) | 4.23 (0.07) | 4.92 | 4.75 (0.01) |
| $S_4$ | 2B (098 $\rightarrow$ 102) | 4.28 (0.04) | 5.05 | 4.79 (0.09) |
| $S_5$ | 3A (098 $\rightarrow$ 103) | 4.49 (0.21) | 5.06 | 5.01 (0.00) |
| | | TPP | | |
| $S_1$ | 1B (101 $\rightarrow$ 103) | 3.77 (0.32) | 4.17 (0.34) | |
| $S_2$ | 1A (101 $\rightarrow$ 102) | 3.78 (0.42) | 4.57 (0.45) | |
| $S_3$ | 2A (099$\rightarrow$ 102) | 4.04 (0.00) | 4.84 (0.00) | |
| $S_4$ | 2B (099 $\rightarrow$ 103) | 4.07 (0.00) | 4.52 (0.00) | |
| $S_5$ | 3B ( 100 $\longrightarrow$ 102 ) | 4.40 (0.03) | 4.92 (0.00) | |

## 3.2 Population dynamics

The diabatic population dynamics reflect the fundamentally different excited-state architectures of TePP and TPP. Since the ground-state population ($S_0$) remains zero throughout the simulation window for both systems (although hops to the ground state are allowed), the dynamics correspond entirely to redistribution within the excited-state manifolds, without any internal conversion to $S_0$ on the sub-200 fs timescale. In TePP, all low-lying diabats possess intramolecular charge-transfer (CT) character emerging from a common pyrrole $\pi$ donor orbital. As a consequence, the initially prepared population undergoes ultrafast intramanifold mixing among CT states that differ only in the localization of the acceptor density on the phenyl substituents. On a sub-10 fs timescale the population spreads from the initially occupied CT state into higher-lying CT diabats, driven by strong vibronic couplings mediated by phenyl torsions and bond-alternation motions along the pyrrole-phenyl junctions. This rapid redistribution persists throughout the simulation: the population remains broadly distributed across multiple CT diabats and does not concentrate in any single state at longer times. The CT manifold of TePP therefore behaves as a dense, strongly coupled network in which electronic character is continuously exchanged among states of similar donor-acceptor topology.

In TPP, the population dynamics reveal a more structured progression through a less strongly coupled manifold. The initial excitation resides in locally excited $\pi\pi^*$ diabats on the pyrazine core, and the early-time dynamics are dominated by population exchange within this LE subspace. Because the LE states are symmetry-stabilized and more energetically isolated at the Franck–Condon geometry, their couplings to the higher-lying $n\pi^*$ diabats grow only gradually as symmetry-breaking distortions develop during the nuclear motion. As a result, population remains confined to the LE manifold for a longer time than in TePP before progressively accessing the $n\pi^*$ states. Once the $n\pi^*$ diabats begin to participate, the redistribution becomes more extensive, but the overall dynamics remain more sequential and less diffusive than in TePP.

Regarding excited-state deactivation timescales, TePP exhibits an ultrafast loss of its initial charge-transfer character, with a half-life of $11 \pm 1$ fs and a $1/e$ decay time of $80 \pm 5$ fs, whereas TPP relaxes more slowly within its $\pi\pi^*$ manifold, displaying a half-life of $33 \pm 2$ fs and a $1/e$ decay time of $105 \pm 5$ fs. The $1/e$ decay time corresponds to the moment when the population of the initially excited diabat has decreased to $1/e \approx 0.37$ of its initial value, while the half-life denotes the time required for the population to fall to 50% of its initial amplitude.

In summary, TePP exhibits fast, widespread mixing within a densely interconnected CT manifold, whereas TPP shows delayed but more stepwise population transfer from LE to $n\pi^*$ character. The absence of any $S_0$ population throughout the dynamics indicates that neither chromophore accesses a productive internal-conversion funnel to the ground state within the first 200 fs, and the observed behavior is therefore governed entirely by excited-state intramanifold couplings driven by their respective electronic structures.

Regarding the role of phenyl groups mentioned in reference 22, an analysis of the dihedrals for TPP and TePP shown in Fig. 5, reveal that, for TPP, there are no significant differences among the four dihedrals between the pyrazine ring and the phenyl groups during the dynamics; in contrast, TePP exhibits a clear difference between the 2,5 and 3,4-phenyl groups (see Figure S1 for atom order). This behavior is aligned with previous studies, setting a different role for each pair of phenyl groups which is shown in the dihedral analysis even in the gas phase. The 2,5-phenyl groups behave in a similar way to the four phenyl groups in TPP, with the angle varying from 40° to values slightly above 20°. In contrast, the 3,4-phenyl groups remain stable, keeping the corresponding dihedral angles between 60-50°. In our previous study, Ref. 41, we also shown that the torsional angle varied slightly for the excited state optimised geometry of TPP.

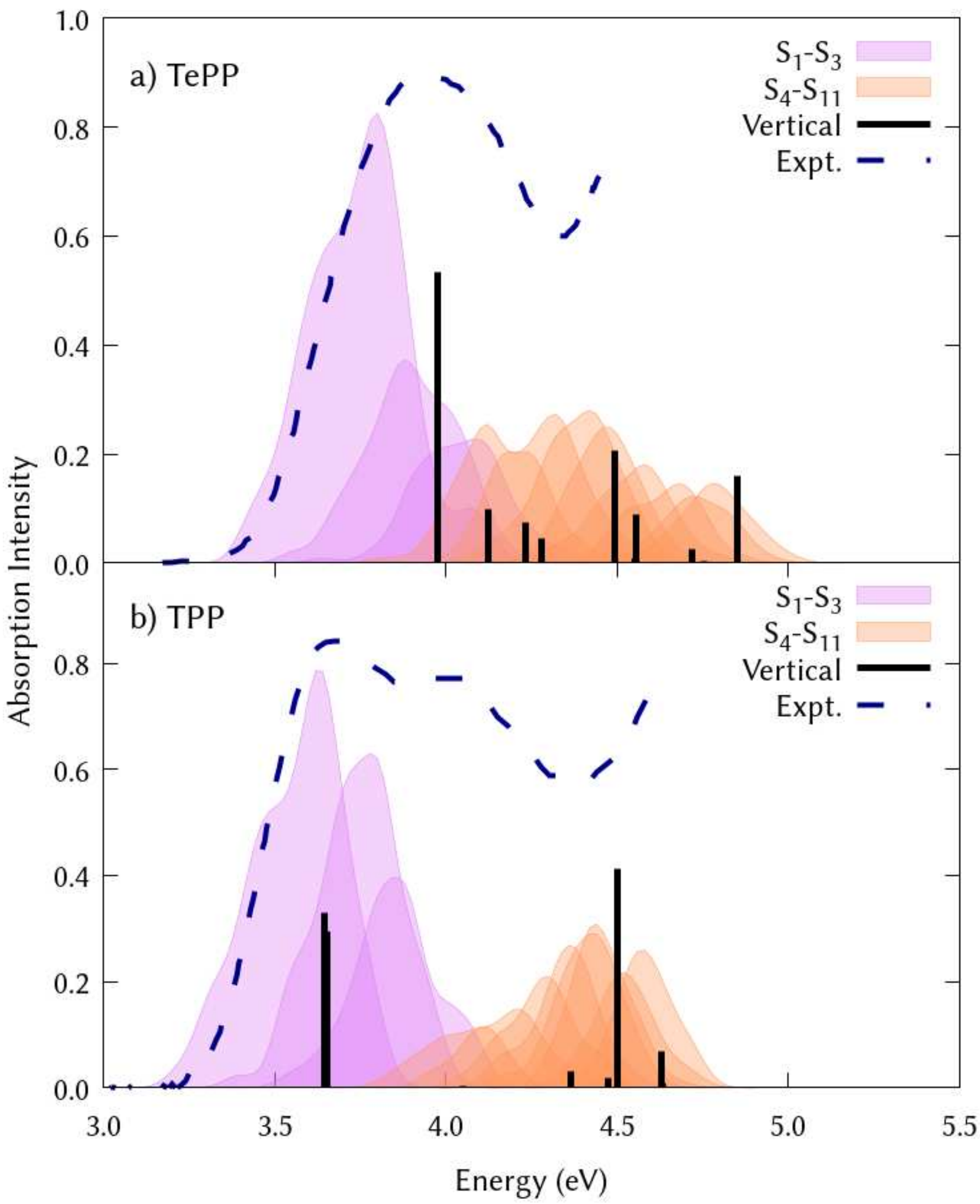

Figure 3: Static absorption spectrum of (a) TePP, (b) TPP computed by vertical excitations from the FC region using TD-DFT (black line) including vibronic coupling by computing vertical excitations from 200 initial conditions selected from a Wigner distribution. Due to their dominant oscillator strengths, the initial photoexcitation primarily populates the $S_1$-$S_3$ adiabatic states, which are highlighted in pink. The experimental absorption spectra are taken from Ref. 22 and Ref. 23.

Further insight into the distinct excited-state relaxation mechanisms of TePP and TPP is obtained by correlating the normal mode analysis with the geometrical characteristics of the hopping events. As shown in Fig. S7, the modes exhibiting the highest dynamical activity in TePP are predominantly low-frequency motions involving phenyl wagging and peripheral N–H distortions, which are very low and very large-amplitude displacements. The associated displacement vectors (Figs. S8) reveal predominantly out-of-plane character, consistent with the broad distributions observed in the hopping geometries projected onto phenyl–phenyl distances and dihedral angles (Fig. S11). This indicates that nonadiabatic transitions in TePP are facilitated by flexible, loosely correlated peripheral motions, enabling efficient population redistribution among emissive states without driving the system toward strongly quenched configurations.

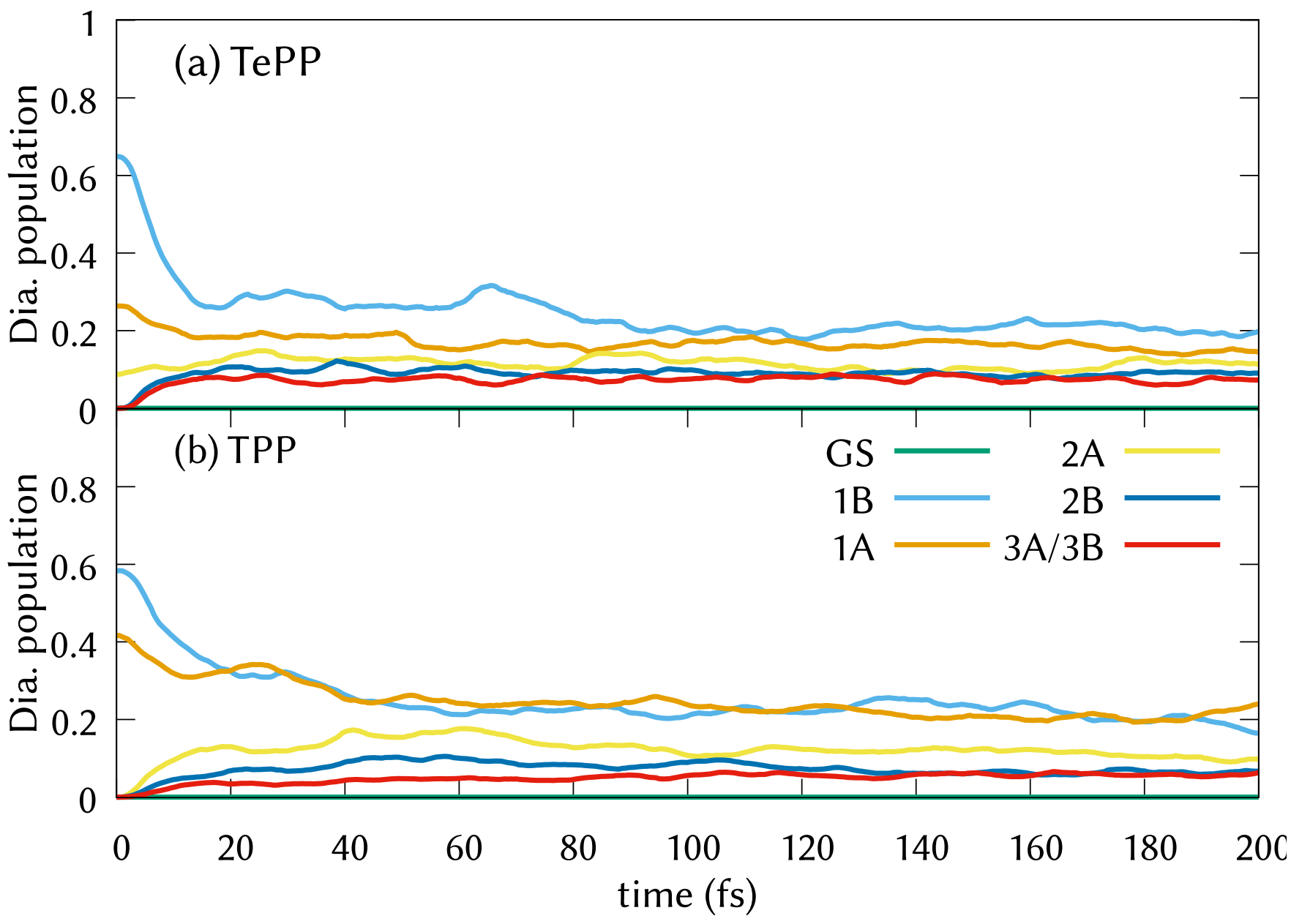

Figure 4: Normalized diabatic populations of TePP (top) and TPP (bottom).The diabatisation has been made tracing back to the state character of the initial populated state via atomic overlaps and populations are collected as the square of the quantum amplitudes of the states [quantum populations]. The $S_5$ state (red line) has A symmetry for TePP and B symmetry for TPP, therefore the 3A/3B label.

In contrast, the normal mode activity of TPP is dominated by mid-frequency modes localized on the central pyrazine core (Fig. S7), involving ring deformations and symmetry-breaking distortions rather than peripheral rotations. The corresponding displacement vectors (Fig. S9) suggest more localized and coordinated nuclear motion, which is reflected in the hopping geometries by modest correlations between central structural distortions and phenyl separations (Fig. S12). These observations indicate that excited-state deactivation in TPP is driven by specific collective distortions of the molecular core that promote access to dark $n\pi^*$ states, rather than by diffuse large-amplitude motion. This points to fundamentally distinct nonradiative relaxation mechanisms: a flexible, weakly correlated pathway in TePP versus a more structured, mode-specific deactivation channel in TPP.

## 3.3 Time-dependent experimental observables

Traditionally, nonadiabatic molecular dynamics has relied heavily on the analysis of electronic state populations to describe excited-state pathways. However, electronic populations are not experimental observables and depend on the electronic representation employed (adiabatic vs. diabatic). Consequently, pronounced changes in the computed populations may correspond to only subtle features in experimentally accessible signals, and vice versa. In line with the distinction between theoretical properties and measurable quantities discussed in our previous work,[42] we focus here on two time-resolved observables that can be computed directly from the nuclear dynamics and linked to ultrafast spectroscopy: transient absorption spectroscopy (TAS) and time-resolved fluorescence (TR-FL). For the two systems investigated, tetraphenylpyrazine (TPP) and tetraphenylpyrrole (TePP), these observables provide a more physically grounded picture of the excited-state dynamics than population-based analyses alone.

For the TR-FL simulations, the time-resolved fluorescence signal was computed directly from the instantaneous emission intensity $A(t)$ extracted from the dynamics, based on

$$A(t) = \omega(t)^3 f_{\text{osc}}(t),$$

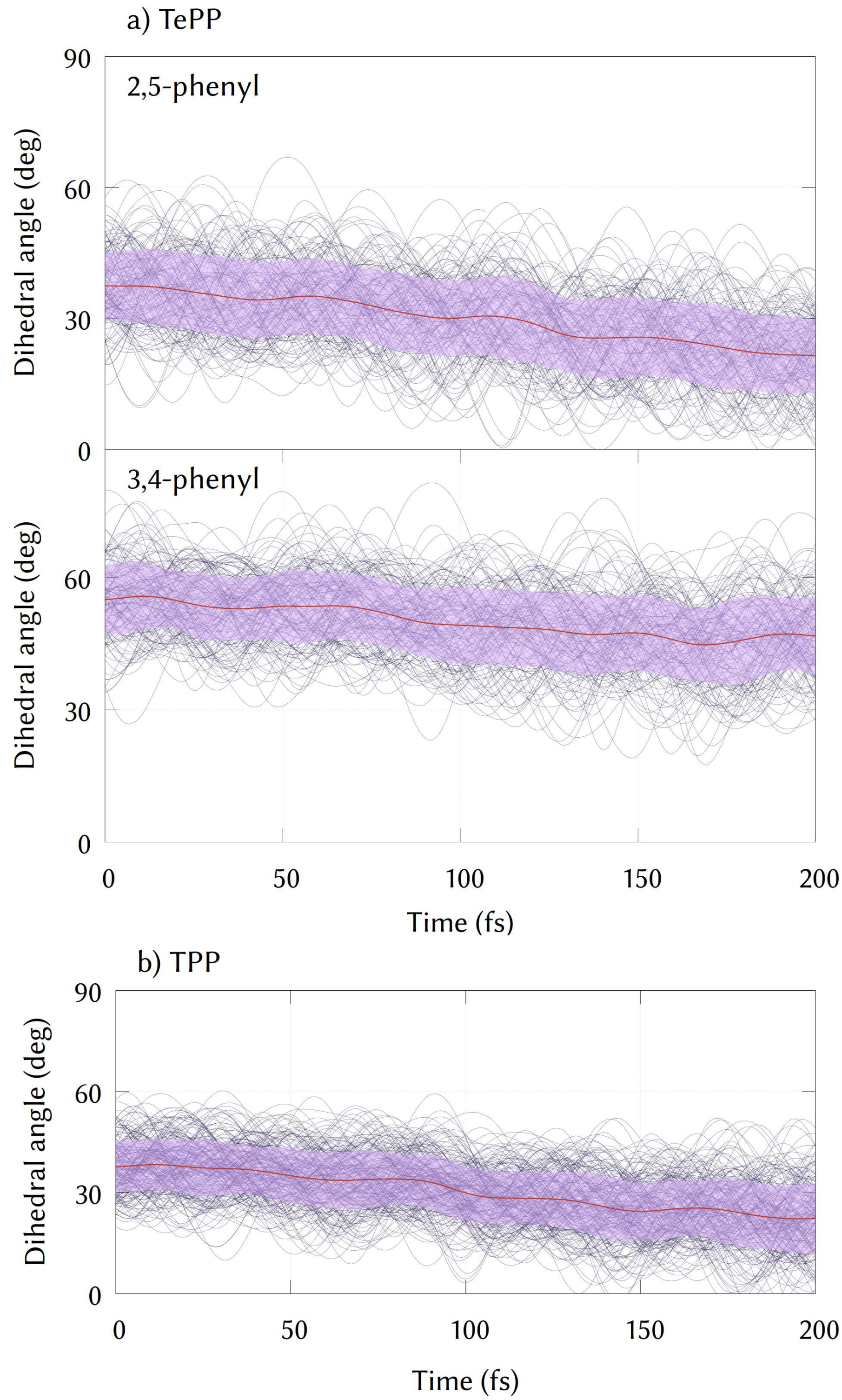

Figure 5: Time evolution of the dihedral angles between the central heterocycle and the phenyl substituents along the trajectories for (a) TePP and (b) TPP. Each line corresponds to a single trajectory; the solid red line represents the ensemble average and the shaded purple region indicates the standard deviation. In TePP, dihedral angles are shown separately for the 2,5-phenyl groups (C2-C1-C18-C20) and the 3,4-phenyl groups (C1-C2-C7-C9). For TPP, only one representative dihedral (C1-C2-C29-C30) is shown due to the equivalence of all four phenyl groups. The time evolution of all dihedral angles is provided in Section S3, Figs. S5-6 of the ESI.

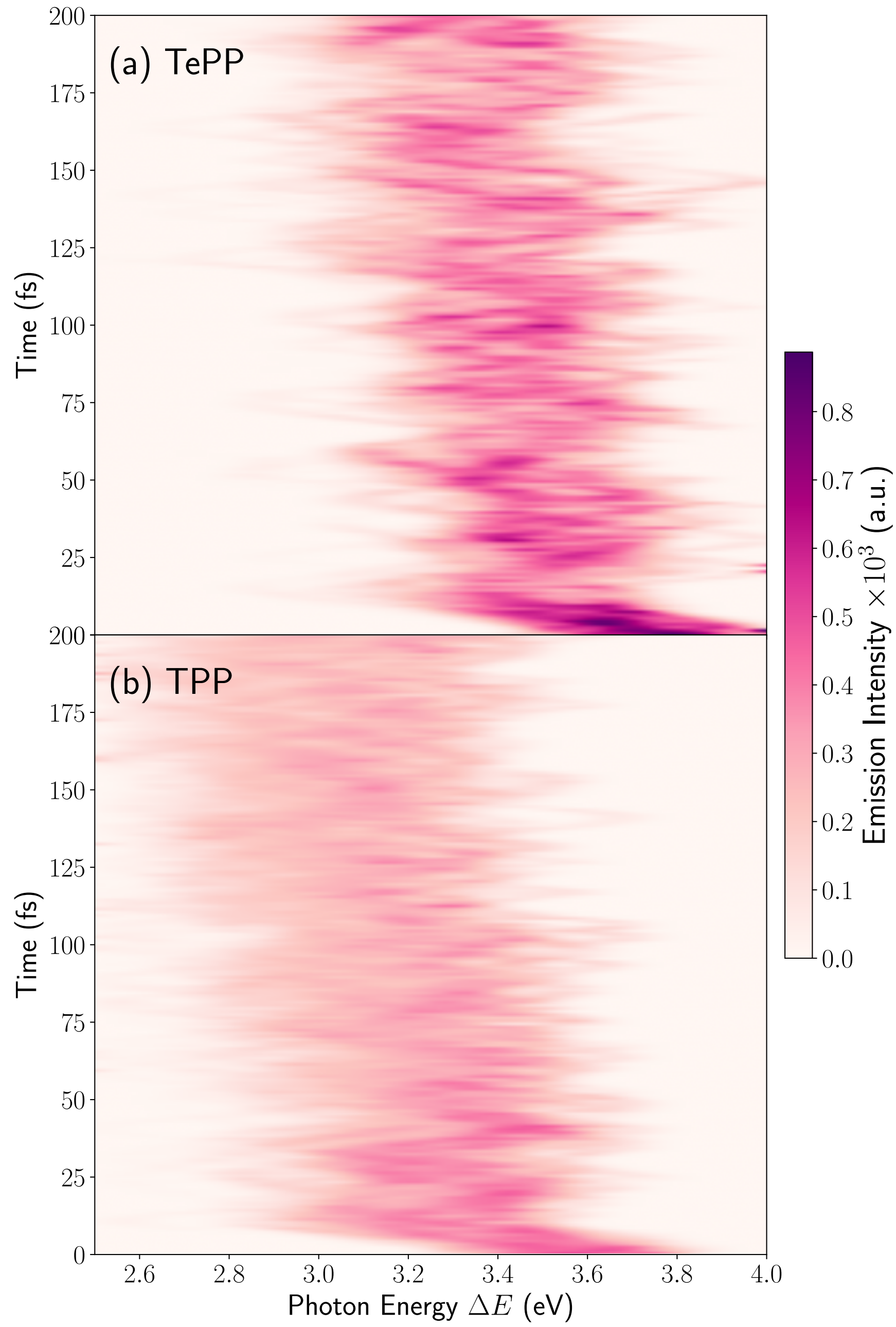

Figure 6: Time-resolved emission maps of (a) TePP and (b) TPP plotted as photon energy $\Delta E$ versus time. The color scale (scaled by $10^3$) reflects the emission intensity derived from surface-hopping trajectories and normalized to TePP. Further details on the construction of these maps are given in the Supporting Information.

retaining only contributions from trajectories that were on the $S_1$ electronic state. This approach provides an observable for comparing the radiative dynamics of TPP and TePP under identical excitation conditions.

The distinct excited-state population dynamics translate into markedly different time-resolved emission behaviors, as shown in Fig. 6. Although population transfer among electronic states is delayed in TPP relative to TePP, the evolution of the emission spectrum exhibits the opposite trend. In TPP [Fig. 6(b)], the emission rapidly shifts to lower photon energies while simultaneously losing intensity, indicating a fast depletion of emissive character. By 200 fs, the emission is broadly distributed over nearly 1 eV (2.6-3.6 eV), consistent with population trapping in the below dark $n\pi^*$ states. In contrast, TePP [Fig. 6(a)] retains significant emissive intensity throughout the entire 200 fs window, with a more modest red shift of approximately 0.5 eV that can be attributed to the gradual stabilization of emissive states driven by nuclear relaxation.

These differences provide a natural explanation for the contrasting photoluminescence behavior of the two chromophores. In solution, TPP acquires sufficient conformational flexibility to efficiently relax into dark $n\pi^*$ states, and likely on longer time scales toward the ground state, resulting in quenched emission. This loss of emission thus appears to be an intrinsic property of the tetraphenylpyrazine molecular framework rather than an aggregation-induced effect such as SLE. In the solid state, however, conformational restriction imposed by molecular packing hinders even more this relaxation pathway, trapping the system in more emissive configurations. TePP, on the other hand, remains emissive in both solution and solid phases, which can be rationalized by the steric constraints imposed by the smaller pyrrole ring: the phenyl substituents are kept in closer proximity, enhancing steric hindrance and stabilizing charge-separated emissive states while inhibiting relaxation into dark manifolds.

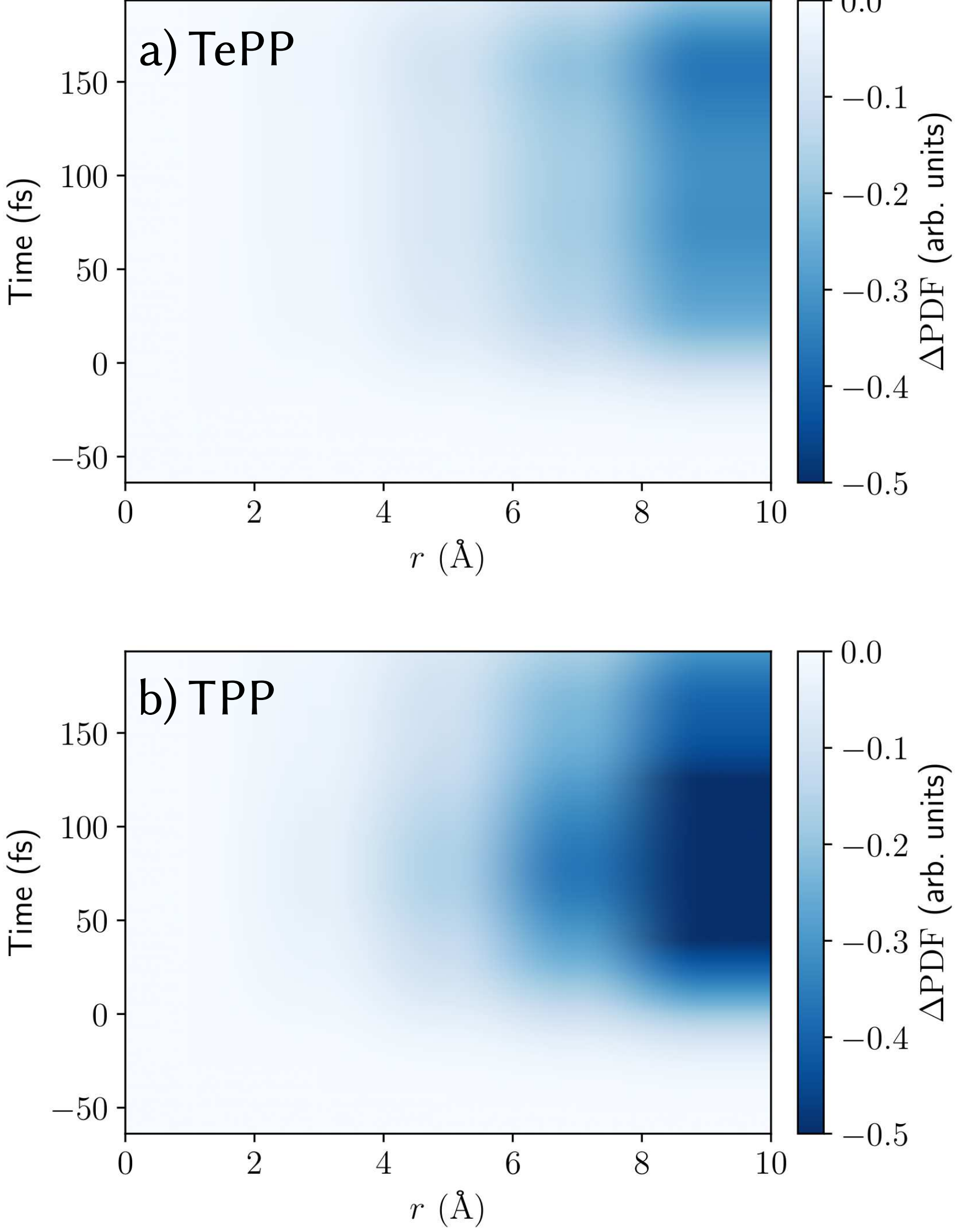

Figure 7: Time-resolved ultrafast electron diffraction (UED) pair distribution function (PDF) signals computed from trajectory surface hopping simulations for (a) TePP and (b) TPP. The PDFs were obtained by averaging over all trajectories and convoluting the scattering signal with gaussians of full width at half maximum (FWHM) of 50 fs. The time evolution reflects structural dynamics following photoexcitation, with a simulation time step of 20 fs up to 200 fs.

We also calculated the ultrafast electron diffraction (GUED) spectrum to investigate the photoinduced structural dynamics of tetraphenylpyrazine (TPP) and tetraphenylpyrrole (TePP). Since no experimental GUED data are currently available for either system, the simulations were performed using parameters guided by previous theoretical and experimental studies of ultrafast electron diffraction in polyatomic organic molecules.[53, 54, 55] The time-resolved scattering signal was computed within the independent atom model and subsequently transformed into real space in order to facilitate a direct structural interpretation.

The time-resolved difference pair distribution function, $\Delta PDF(r,t)$, is defined as the difference between the instantaneous pair distribution function $P(r,t)$ and the reference distribution at time zero,

$$\Delta PDF(r,t) = PDF(r,t) - PDF(r,0), \tag{2}$$

where the reference corresponds to the first frame of the nuclear dynamics. The PDF was obtained from the molecular scattering signal via a sine Fourier transform with an exponential damping factor to suppress high-$Q$ contributions. The resulting $\Delta PDF(r,t)$ signal was evaluated over the range $r = 0$–$10$ Å and convoluted along the time axis with a Gaussian instrument response function to account for the finite temporal resolution of GUED experiments. More details are given in section S4 of the Supporting Information.

The time-resolved ΔPDF spectra display distinct spatial and temporal characteristics following photoexcitation. For TPP, the ΔPDF signal stronger overall, with negative features extending to large interatomic distances ($r \gtrsim$ 8-9 Å) and persisting over the entire simulation time, reflecting sustained structural distortions of both the pyrazine core and the peripheral phenyl groups In contrast, TePP exhibits weaker ΔPDF response, approximately half the magnitude of TPP, with negative features at large distances ($r \gtrsim$ 8-9 Å) that are most pronounced at early times (50–100 fs) and partially subside at longer times. The larger amplitude of the TPP signal suggests a more pronounced structural distortion upon photoexcitation, consistent with the dominant role of pyrazine core distortions in driving relaxation away from the bright $\pi\pi^*$ state.

# 4 Conclusions

In this article, we aimed to rationalize the markedly different emissive properties of TePP and TPP. As shown in Fig. 1, TePP exhibits a high photoluminescence quantum yield both in THF solution and in the crystalline phase, indicating that its radiative decay pathways remain efficient across different environments. TPP displays much weaker emission overall, with a quantum yield in solution that is approximately one order of magnitude lower than in the solid state and substantially reduced relative to TePP. Given that THF is an apolar solvent, we hypothesized that excited-state relaxation pathways observed in the gas phase could already capture essential aspects of the solution-phase behavior, provided that nonradiative deactivation is dominated by intrinsic molecular mechanisms rather than by strong solvent-specific effects. To test this hypothesis, we performed gas-phase trajectory surface hopping simulations using the same electronic-structure methodology that has previously been shown to reliably reproduce the absorption and emission spectra of TPP.[41]

As a result, the comparative excited-state dynamics of TePP and TPP reveal that their contrasting photoluminescence behavior originates from qualitatively different intramolecular relaxation mechanisms rather than from a single environmental effect. For TePP, the dynamics are characterized by rapid redistribution of population among several electronically mixed excited states, while maintaining access to emissive configurations over the full simulation time. This behavior is reflected in the approximately constant emission intensity and gradual red shift observed in the time-resolved emission map (Fig. 6a), consistent with nuclear relaxation on the excited-state surfaces. This picture is further supported by the simulated GUED spectra, which show ΔPDF features that peak at early times (50–100 fs) at large interatomic distances ($r > 6$ Å), consistent with a transient correlated displacement of the peripheral phenyl groups that partially subsides at longer times. The normal mode analysis indicates that low-frequency peripheral motions, such as phenyl wagging and N-H distortions, are most actively involved in the dynamics, suggesting relaxation pathways that are compatible with emission in both solution and solid phases.

By contrast, TPP exhibits a markedly different evolution. Although ground-state quenching is not observed in the gas-phase dynamics, the calculated time-resolved emission map (Fig. 6b) shows a pronounced loss of intensity accompanied by a red shift, indicating population transfer toward weakly emissive or dark excited states. Consistently, the simulated GUED response of TPP is dominated by ΔPDF features of approximately twice the magnitude of TePP, with signal extending to large interatomic distances ($r > 6$ Å) that persists over the entire simulation time, reflecting both stronger structural distortions of the pyrazine core and sustained correlated displacement of the peripheral phenyl groups. The normal mode analysis identifies central pyrazine distortions as the dominant modes driving this relaxation, pointing to an intrinsic intramolecular pathway that funnels population away from bright $\pi\pi^*$ states. These observations suggest that the lack of luminescence of TPP in solution may already be largely encoded in its intrinsic excited-state relaxation pathways. The gas-phase dynamics capture both the progressive red shift of the emission and a pronounced loss of intensity even in the absence of environment. Since the dominant nuclear motions driving this relaxation involve central pyrazine distortions,which are expected to persist in both gas phase and solution, and since dark $n\pi^*$ states are not anticipated to be strongly destabilized in THF (nonpolar solvent), the gas-phase simulations likely provide a realistic qualitative description of the solution-phase dynamics. In this context, solvent effects are expected to play a secondary role, modulating excited-state energetics rather than introducing fundamentally new deactivation channels. By contrast, the enhanced emission observed in the solid state is more naturally attributed to environmental constraints that hinder access to dark-state trapping regions or reduce vibronic coupling efficiency, rather than to a complete alteration of the underlying intramolecular relaxation mechanism.

**Supporting Information**

All data necessary to reproduce the results presented in this article are available at Zenodo at `10.5281/zenodo.19639017`. The dataset includes the velocities and geometries from the wigner distribution `initconds`, trajectory geometries, the vibrational frequency calculations provided as Molden files and the Supporting Information as a PDF file, which contains reference geometries of TePP and TPP (Section S1), convergence tests with respect to the trajectory ensemble size (Section S2), analysis of dihedral angle evolution (Section S3), additional details on plot construction (Section S4), normal mode analysis (Section S5), and representative hopping geometries (Section S6).

**Acknowledgements**

SG thanks EPSRC under the COSMOS programme grant (EP/X026973/1) and the Spanish Ministry of Science and Innovation grant (PID2024-162002NB-I00). J.H.R, A.M.S and S.G.C acknowledge funding by Spanish Ministry of Science and Innovation (MCIN/AEI/10.13039/501100011033) grant No. PID2020-113147GA-I00 and PID2023-147215NB-I00 funded by Grant No. MCIN/AEI/10.13039/501100011033 Spanish Ministry of Science and Innovation, and FEDER, UE. SGC acknowledge funding by the Spanish Ministry of Science and Innovation (Grants No. PID2021-122549NB-C21 and PID2024-156686NB-I00). This research has made use of the high performance computing resources of the Castilla y León Supercomputing Center (SCAYLE), financed by the European Regional Development Fund (ERDF). The authors thank the COST action CA21101" Confined molecular systems: from a new generation of materials to the stars (COSY)" supported by COST (European Cooperation in Science and Technology) for inspiring discussions and enabling cooperative work.